\documentclass[preprints,article,accept,moreauthors,
pdftex,10pt,a4paper]{mdpi} 

\firstpage{1} 
\articlenumber{x}
\doinum{10.3390/------}
\pubvolume{2}
\pubyear{2018}
\copyrightyear{2018}
\externaleditor{Academic Editor: Serghei Klimin}
\history{}

\usepackage[english]{babel}
\usepackage[utf8]{inputenc}
\usepackage{graphicx}
\usepackage{siunitx}
\usepackage{amsmath}
\usepackage{amsfonts}
\usepackage{amssymb}
\usepackage{graphicx}

\newcommand{\beq}{\begin{equation}}
\newcommand{\eeq}{\end{equation}}
\newcommand{\beqa}{\begin{eqnarray}}
\newcommand{\eeqa}{\end{eqnarray}}

\Title{Self-consistent derivation of the modified Gross-Pitaevskii 
equation with Lee-Huang-Yang correction} 

\Author{L. Salasnich$^{1,2}$} 

\address{$^{1}$Dipartimento di Fisica e Astronomia ``Galileo Galilei'' and 
CNISM, Universit\`a di Padova, Via Marzolo 8, 35131 Padova, Italy \\
$^{2}$Istituto Nazionale di Ottica (INO) del Consiglio Nazionale 
delle Ricerche (CNR), Via Nello Carrara 1, 50019 Sesto Fiorentino, Italy}

\corres{Correspondence: luca.salasnich@unipd.it}

\abstract{We consider a dilute and ultracold bosonic gas of weakly-interacting 
atoms. Within the framework of quantum 
field theory we derive a zero-temperature modified 
Gross-Pitaevskii equation with beyond-mean-field corrections  
due to quantum depletion and anomalous density. 
This result is obtained from the stationary equation 
of the Bose-Einstein order parameter coupled 
to the Bogoliubov-de Gennes equations of 
the out-of-condensate field operator. We show that, 
in the presence of a generic external trapping potential, 
the key steps to get the modified Gross-Pitaevskii equation 
are the semiclassical approximation for the Bogoliubov-de Gennes equations, 
a slowly-varying order parameter, and a small quantum depletion. 
In the uniform case, from the modified Gross-Pitaevskii equation 
we get the familiar equation of state with Lee-Huang-Yang correction.} 


\begin{document}

\section{Introduction}

In 1924 Bose and Einstein introduced the concept 
of Bose-Einstein statistics and also of Bose-Einstein condensation, i.e. 
the macroscopic occupation of the lowest single-particle state of 
a system of bosons \cite{bose,einstein}. 
In 1938 London suggested that the normal-superfluid 
phase transition of $^4$He is related to the 
Bose-Einstein condensation and to the existence of a 
macroscopic wavefunction for the Bose condensate \cite{london1,london2}. 
In 1947 Bogoliubov calculated, for a uniform weakly-interacting Bose gas, 
the quantum depletion, i.e the fraction of 
bosons which are not in the Bose condensate at zero temperature 
due to a repulsive interaction strength \cite{bogoliubov}. 
In 1957 Lee, Huang and Yang evaluated the first correcting term 
to the mean-field equation of state of a uniform and weakly-interacting 
Bose gas \cite{lhy1957}. In 1961 Gross and Pitaevskii 
derived the mean-field equation for the space-dependent 
macroscopic wavefunction of a weakly-interacting Bose gas 
in the presence of an external trapping potential \cite{gross,pitaevskii}. 
The Gross-Pitaevskii equation is the main tool used to describe 
the properties of the Bose-Einstein condensates which are now 
routinely produced with ultracold and dilute 
alkali-metal atoms \cite{leggett}. 

Some years ago, experiments with atomic gases 
reported evidence of beyond-mean-field effects on the 
equation of state of repulsive bosons 
\cite{papp2008,wild2012}. These experimental results 
are quite well reproduced \cite{sala-zero} by a modified Gross-Pitaevskii 
equation which includes a beyond-mean-field correction 
that is the local version of the Lee-Huang-Yang term. 
Few years ago, Petrov suggested theoretically the existence of
self-bound quantum droplets in an attractive Bose-Bose mixture,
where the collapse is suppressed by a beyond-mean-field term \cite{petrov}. 
Very recent experiments \cite{cabrera,semeghini} 
with two internal states of $^{39}$K atoms in a 
three-dimensional configuration substantially confirm 
these theoretical predictions based on a modified Gross-Pitaevskii equation.

Beyond-mean-field correcting terms into the Gross-Pitaevskii 
equation \cite{polls1,sala1,polls2}, or into similar 
nonlinear Schr\"odinger equations for superfluids 
\cite{sala2,sala3}, are usually introduced heuristically 
in the spirit of the density functional theory. 
Here we derive the modified Gross-Pitaevskii equation 
in a self-consistent way starting from the Heisenberg equation 
of motion of the bosonic field operator ${\hat \psi}({\bf r},t)$ 
and the familiar Bogolibov prescription of writing the quantum field operator 
as the sum of a classical complex field $\psi_0({\bf r})$, 
that is the order parameter or macroscopic wavefunction of 
the Bose-Einstein condensate, and a quantum field ${\hat \eta}({\bf r},t)$ 
which takes into account quantum and thermal 
fluctuations \cite{stoof,griffin}. The presence of a generic external trapping 
potential $U({\bf r})$ is circumvented by adopting 
a semiclassical approximation 
for the Bogoliubov-de Gennes equations of the fluctuating quantum 
field \cite{giorgini,giorgini1}. In this way, at zero temperature 
we obtain the local density 
${\tilde n}({\bf r})$ of the out-of-condensate bosons as a function 
of the classical field $\psi_0({\bf r})$ and the corresponding 
equation for $\psi_0({\bf r})$, that is the stationary modified 
Gross-Pitaevskii equation with beyond-mean-field terms. From these 
terms we recover the Lee-Huang-Yang correction \cite{lhy1957} 
in the case of a uniform and real Bose-Einstein order parameter. 

\section{Quantum field theory of bosons} 

Let us consider the bosonic quantum field 
operator ${\hat \psi}({\bf r},t)$ describing 
a non-relativistic system of confined and interacting 
identical atoms in the same hyperfine state. Its 
Heisenberg equation of motion is given by \cite{stoof}
\beq 
i\hbar {\partial \over \partial t} {\hat \psi}({\bf r},t) =
\Big[ -{\hbar^2\over 2m} \nabla^2 
+ U({\bf r}) - \mu \Big] {\hat \psi} ({\bf r},t)
+ g \ {\hat \psi}^+({\bf r},t)
{\hat \psi}({\bf r},t){\hat \psi}({\bf r},t) \; .  
\label{heq}
\eeq
where $m$ is the mass of the atom, 
$U({\bf r})$ is the confining external potential,  
$g$ is the strength of the interatomic potential, 
and $\mu$ is the chemical potential, which is fixed 
by the conservation of the particle number $N$, that is 
an eigenvalue of the number operator 
\beq 
{\hat N}= \int d^3 {\bf r} \; 
{\hat \psi}^+({\bf r},t){\hat \psi}({\bf r},t) \; . 
\eeq 
The bosonic field operator  ${\hat \psi}({\bf r},t)$ satisfies the familiar 
equal-time commutation rules. In Eq. (\ref{heq}) we have assumed 
that the system is very dilute 
and such that the scattering length and the range of the 
interatomic interaction are much smaller than the average interatomic 
distance. Thus, the true interatomic potential is 
approximated by a local pseudo-potential 
\beq 
V({\bf r},{\bf r}')=g \ \delta^3({\bf r}-{\bf r}') \; , 
\eeq
where 
\beq 
g={4\pi \hbar^2 a_s\over m} 
\eeq 
is the scattering amplitude of the spin triplet channel 
with $a_s$ the s-wave scattering length \cite{stoof}.

\section{Bogoliubov prescripition and quantum fluctuations}

In a bosonic system one can separate Bose-condensed 
particles from non-condensed ones by using 
of Bogoliubov prescription \cite{stoof,griffin} 
\beq 
{\hat \psi}({\bf r},t)=\psi_0({\bf r}) 
+{\hat \eta}({\bf r},t) \; , 
\eeq 
where 
\beq 
\psi_0({\bf r}) =\langle {\hat \psi}({\bf r},t) \rangle 
\eeq 
is the time-independent but space-dependent 
complex order parameter (macroscopic wavefunction)
of the Bose-Einstein condensate with $\langle ... \rangle$ 
the thermal average over an equilibrium ensemble. 
Notice that we work at thermal equilibrium 
and consequently the thermal averages are time independent. 
The field ${\hat \eta}({\bf r},t)$ is the operator of quantum and 
thermal fluctuations, which describes out-of-condensate bosons. 

The Bogoliubov prescription for the field operator 
${\hat \psi}({\bf r},t)$ enables us to write 
the three-body thermal average in the following way 
\beq 
\langle 
{\hat \psi}^+({\bf r},t) 
{\hat \psi}({\bf r},t){\hat \psi}({\bf r},t)
\rangle = 
|\psi_0({\bf r})|^2 \psi_0({\bf r}) 
+ 
2 \tilde{n}({\bf r}) \ \psi_0({\bf r}) 
+ \tilde{m}({\bf r}) \ \psi_0^*({\bf r}) 
+ \tilde{s}({\bf r})  \; ,
\eeq
where $\tilde{n}({\bf r})=\langle 
{\hat \eta}^+({\bf r},t){\hat \eta}({\bf r},t)\rangle$ 
is the density of non-condensed particles, while $\tilde{m}({\bf r})=\langle 
{\hat \eta}({\bf r},t){\hat \eta}({\bf r},t)\rangle$ 
is the anomalous density and $\tilde{s}({\bf r})=\langle 
{\hat \eta}^+({\bf r},t){\hat \eta}({\bf r},t)
{\hat \eta}({\bf r},t)\rangle$ is the anomalous correlation \cite{griffin}. 

Now, we obtain an equation for $\psi_0({\bf r})$ 
by taking the thermal average on Eq. (\ref{heq}). 
In this way we find 
\beq 
\mu \ \psi_0({\bf r}) = 
\Big[ -{\hbar^2\over 2m} \nabla^2 + U({\bf r}) 
+ g |\psi_0({\bf r})|^2 + 2 g \tilde{n}({\bf r}) \Big] 
\psi_0({\bf r}) 
+ g \tilde{m}({\bf r})\psi_0^*({\bf r}) + g 
\tilde{s}({\bf r})  \; , 
\label{exact}
\eeq 
that is the exact equation of motion of the Bose-Einstein 
order parameter $\psi_0({\bf r})$ \cite{griffin,giorgini}. 
This is not a closed equation due to the presence 
of the non-condensed density $\tilde{n}({\bf r})$ 
and of the anomalous densities $\tilde{m}({\bf r})$ and 
$\tilde{s}({\bf r})$. Neglecting the non-condensed density 
and the anomalous densities the previous equation becomes 
\beq 
\mu \ \psi_0({\bf r})
= \Big[ -{\hbar^2\over 2m} \nabla^2 
+ U({\bf r}) + g |\psi_0({\bf r})|^2 \Big] \psi_0({\bf r}) \; , 
\eeq 
that is the familiar Gross-Pitaevskii equation \cite{gross,pitaevskii}. 
A less drastic approximation, that is called 
Bogoliubov-Popov-Beliaev approximation \cite{griffin,giorgini,giorgini1}, 
neglects only the term $\tilde{s}({\bf r})$. 
Then the equation of motion of the Bose-Einstein order 
parameter $\psi_0({\bf r})$ becomes 
\beq 
\mu \ \psi_0({\bf r}) = 
\Big[ -{\hbar^2\over 2m} \nabla^2 
+ U({\bf r})  + 
g |\psi_0({\bf r})|^2 + 2 g \tilde{n}({\bf r}) \Big] 
\psi_0({\bf r}) + g \tilde{m}({\bf r}) \psi_0^*({\bf r}) \; . 
\label{ggpe}
\eeq 
Also this equation is not closed. We must add an equation 
for the non-condensed density $\tilde{n}({\bf r})$ and the 
anomalous density $\tilde{n}({\bf r})$ 
by studying the fluctuation operator ${\hat \eta}({\bf r},t)$ 
\cite{griffin,giorgini}. 

The equation of motion of the 
fluctuation operator ${\hat \eta}({\bf r},t)$ is obtained 
by subtracting Eq. (\ref{ggpe}) from Eq. (\ref{heq}). 
The standard Bogoliubov-Popov approximation \cite{griffin,giorgini}
neglects both the non-condensate density and the anomalous terms, 
and it takes only linear terms of ${\hat \eta}({\bf r},t)$ and 
${\hat \eta}^+({\bf r},t)$. In this way 
the linearized equation of motion of the fluctuation operator reads 
\beq 
i\hbar {\partial \over \partial t} {\hat \eta}({\bf r},t) =
\Big[ -{\hbar^2\over 2m} \nabla^2 + U({\bf r}) - \mu 
+ 2 g |\psi_0({\bf r})|^2 \Big] {\hat \eta}({\bf r},t) 
+ g \psi_0({\bf r})^2 {\hat \eta}^+({\bf r},t) \; .  
\label{linearized}
\eeq 

\section{Bogoliubov-de Gennes equations 
and their semiclassical approximation}

The fluctuation operator can be written as 
\beq 
{\hat \eta}({\bf r},t)= \sum_j \Big[ u_j({\bf r})
e^{-iE_jt/\hbar}{\hat a}_j +v_j({\bf r})
e^{iE_jt/\hbar}{\hat a}_j^+ \Big] \; , 
\label{expand}
\eeq 
where ${\hat a}_j$ and ${\hat a}_j^+$ are bosonic operators and 
the real functions $u_{j}({\bf r})$ and $v_{j}({\bf r})$ 
are the wavefunctions of the 
quasi-particle and quasi-hole excitations of energy $E_j$ \cite{giorgini}. 
As a consequence one finds 
\beq 
{\tilde n}({\bf r})= \sum_j \Big[ 
\Big( u_j({\bf r})^2+v_j({\bf r})^2
\Big ) \langle {\hat a}^+_j {\hat a}_j \rangle + v_j({\bf r})^2 
\Big] \; ,
\eeq
and
\beq
\langle {\hat a}^+_j {\hat a}_j \rangle =
{1\over e^{E_j/k_B T}-1}
\eeq
is the Bose factor at temperature $T$ with $k_B$ the Boltzmann constant. 
We stress that at zero temperature one gets 
\beq 
{\tilde n}({\bf r}) = \sum_j v_j({\bf r})^2 \; ,  
\eeq 
and also 
\beq 
{\tilde m}({\bf r}) = \sum_j u_j({\bf r}) \ v_j({\bf r}) \; . 
\eeq

By inserting Eq. (\ref{expand}) into Eq. (\ref{linearized}) 
we obtain the Bogoliubov-de Gennes equations
\beqa
{\hat L} u_j({\bf r}) + g \psi_0({\bf r})^2 
v_j({\bf r}) &=& E_j u_j({\bf r}) \; ,
\\
{\hat L} v_j({\bf r})+g \psi_0^*({\bf r})^2 
u_j({\bf r}) &=& - E_j v_j({\bf r}) \; .  
\eeqa
where 
\beq 
{\hat L} = -{\hbar^2\over 2m} \nabla^2+ U({\bf r}) -\mu
+ 2 g |\psi_0({\bf r})|^2  \; . 
\eeq
The solution of these equation can be done numerically by 
chosing the external potential $U({\bf r})$. 
However, an analytical solution can be obtained within the 
semiclassical approximation, where 
$ -i{\boldsymbol \nabla} \to {\bf k}$ and 
$\sum_j \to \int d^3{\bf k}/(2\pi)^3$ \cite{giorgini}. It follows 
that the Bogoliubov differential equations become algebraic equations 
\beqa
L_{\bf k}({\bf r}) \ u_{\bf k}({\bf r}) 
+g \psi_0({\bf r})^2
v_{\bf k}({\bf r}) &=& E_{\bf k}({\bf r}) \ u_{\bf k}({\bf r}) \; ,
\\
L_{\bf k}({\bf r}) \ v_{\bf k}({\bf r}) + g \psi_0^*({\bf r})^2
u_{\bf k}({\bf r}) &=& - E_{\bf k}({\bf r}) \ v_{\bf k}({\bf r}) \; , 
\eeqa
where 
\beq 
L_{\bf k}({\bf r}) = {\hbar^2k^2\over 2m} + U({\bf r}) -\mu
+ 2 g |\psi_0({\bf r})|^2 \; , 
\eeq
and that the zero-temperature non-condensed density reads 
\beq
{\tilde n}({\bf r}) = \int {d^3{\bf k}\over (2\pi)^3} 
v_{\bf k}({\bf r})^2 \; .
\label{soloio}
\eeq
This quantity is also called local quantum depletion of 
the Bose-Einstein condensate. In addition, the local anomalous 
density is given by 
\beq 
{\tilde m}({\bf r}) = \int {d^3{\bf k}\over (2\pi)^3} 
u_{\bf k}({\bf r}) \ v_{\bf k}({\bf r})  \; .
\label{solotu}
\eeq

\section{Local quantum depletion and generalized 
Gross-Pitaevskii equation}

Assuming a slowly-varying order parameter, such that 
the gradient term can be neglected, but also a small quantum depletion, 
from Eq. (\ref{ggpe}) the chemical potential $\mu$ can be approximated as 
\beq 
\mu \simeq  g |\psi_0({\bf r})|^2 + U({\bf r}) \; . 
\eeq
It is then straightforward to derive the elementary excitations 
\beq 
E_{\bf k}({\bf r}) = \sqrt{{\hbar^2k^2\over 2m}\left({\hbar^2k^2\over 2m} 
+ 2 g |\psi_0({\bf r})|^2 \right)} \; , 
\eeq
and the real quasi-particle amplitudes 
\beqa 
u_{\bf k}({\bf r}) &=& {1\over \sqrt{2}} 
\left( {{\hbar^2k^2\over 2m} +  g |\psi_0({\bf r})|^2 
\over E_{\bf k}({\bf r})} + 1 \right)^{1/2}
\label{u}
\\
v_{\bf k}({\bf r}) &=& - {1\over \sqrt{2}} 
\left( {{\hbar^2k^2\over 2m} +  g |\psi_0({\bf r})|^2 
\over E_{\bf k}({\bf r})} - 1 \right)^{1/2} \; . 
\label{v}
\eeqa
We can now insert Eq. (\ref{v}) into (\ref{soloio}) 
and after integration over the linear momenta we obtain 
\beq 
{\tilde n}({\bf r}) = {\sqrt{2}\over 12\pi^2} 
\left({2 m g\over \hbar^2} \right)^{3/2} |\psi_0({\bf r})|^3 \; . 
\eeq
This is the local version of the familiar Bogoliubov term for 
the quantum depletion, originally obtained for a uniform 
bosonic system, i.e. with $U({\bf r})=0$. 
For the local anomalous average density, after dimensional 
regularization \cite{sala-zero}, we find instead 
\beq 
\tilde{m}({\bf r}) = 3 \ \tilde{n}({\bf r}) \; . 
\eeq

Finally, inserting this expression into Eq. (\ref{ggpe}) we get 
\beqa 
\mu \ \psi_0({\bf r}) &=& 
\Big[ -{\hbar^2\over 2m} \nabla^2 
+ U({\bf r})  + g |\psi_0({\bf r})|^2 
+ {\sqrt{2}\over 6\pi^2} 
\left({2 m\over \hbar^2} \right)^{3/2} g^{5/2} |\psi_0({\bf r})|^3 
\Big] \psi_0({\bf r}) 
\nonumber 
\\
&+& {\sqrt{2}\over 4\pi^2} 
\left({2 m g\over \hbar^2} \right)^{3/2} g^{5/2} 
|\psi_0({\bf r})|^3 \psi_0^*({\bf r})  \; . 
\label{gpeboh}
\eeqa
This is a modified Gross-Pitaevskii equation containing 
beyond-mean-field corrections due the the presence of local 
quantum depletion and anomalous average density. The chemical 
potential $\mu$ of Eq. (\ref{gpeboh}) is fixed by the 
normalization condition 
\beq 
N = \int d^3{\bf r} \left[ |\psi_0({\bf r})|^2 + {\tilde n}({\bf r}) \right]  
= \int d^3{\bf r} \left[ |\psi_0({\bf r})|^2 + 
{\sqrt{2}\over 12\pi^2} 
\left({2 m g\over \hbar^2} \right)^{3/2} |\psi_0({\bf r})|^3 \right] 
\eeq
with $N$ the total number of bosons. 

It is important to stress that, in the case of a uniform 
and real Bose-Einstein condensate, Eq. (\ref{gpeboh}) gives 
\beq 
\mu = g n_0 + 2 g {\tilde n} + 3 g {\tilde n} = g n + 4 g {\tilde n} = 
g n + g n_0 {32\over 3\sqrt{\pi}} \sqrt{n_0 a_s^3} \; ,  
\eeq
which is the chemical potential 
with the familiar beyond-mean-field Lee-Huang-Yang correction 
\cite{lhy1957} under the assumption of small quantum depletion. 
Clearly, one does not obtain this zero-temperature result 
neglecting the anomalous average density ${\tilde m}$. 

\section{Conclusions}

We have shown that a modified Gross-Pitaevskii equation with 
local beyond-mean-field terms can be obtained in a straightforward way 
from a quantum-field-theory formulation without invoking the density 
functional theory. However, the derivation is not exact 
because one performs some approximations on the spectrum of elementary 
excitations and on the spatial dependence of the macroscopic 
wavefunction of the Bose-Einstein condensate. In Ref. \cite{stringari} 
it is shown a different derivation of a zero-temperature 
stationary modified Gross-Pitaevskii equation without external 
confinement but which also takes into account anomalous averages. 
A more formal and mathematical approach 
to the Hartree-Fock-Bogoliubov methods to obtain 
time-dependent modified Gross-Pitaevskii equations 
can be found in Ref. \cite{math}. 
In the next future we want to derive and use coupled 
modified Gross-Pitaevskii equations for studying Bose-Bose mixtures 
under double-well confinement and spin-orbit coupling, 
extending our previous results \cite{sala-double,sala-soc}. 

\section*{Acknowledgements}

The author thanks A. Cappellaro, 
A. Recati, S. Stringari, F. Toigo, A. Tononi, 
and Z.Q. Yu for useful discussions and suggestions. 
The author acknowledges for partial support the FFABR grant of 
Italian Ministry of Education, University and Research. 

\conflictsofinterest{The author declares no conflict of interest.}

\reftitle{References}

\end{document}